\documentclass[preprint]{aastex}
\tighten
\eqsecnum

\begin{document}

\font\twelvei = cmmi10 scaled\magstep1 
       \font\teni = cmmi10 \font\seveni = cmmi7
\font\mbf = cmmib10 scaled\magstep1
       \font\mbfs = cmmib10 \font\mbfss = cmmib10 scaled 833
\font\msybf = cmbsy10 scaled\magstep1
       \font\msybfs = cmbsy10 \font\msybfss = cmbsy10 scaled 833
\textfont1 = \twelvei
       \scriptfont1 = \twelvei \scriptscriptfont1 = \teni
       \def\mit{\fam1 }
\textfont9 = \mbf
       \scriptfont9 = \mbfs \scriptscriptfont9 = \mbfss
       \def\bmit{\fam9 }
\textfont10 = \msybf
       \scriptfont10 = \msybfs \scriptscriptfont10 = \msybfss
       \def\bmsy{\fam10 }

\def\etal{{\it et al.~}}
\def\eg{{\it e.g.,~}}
\def\ie{{\it i.e.,~}}
\def\lsim{\raise0.3ex\hbox{$<$}\kern-0.75em{\lower0.65ex\hbox{$\sim$}}} 
\def\gsim{\raise0.3ex\hbox{$>$}\kern-0.75em{\lower0.65ex\hbox{$\sim$}}} 
\def\kms{~{\rm km~s^{-1}}}
\def\cm3{~{\rm cm^{-3}}}
\def\yr{~{\rm yr}}
\def\Msun{~{\rm M}_{\sun}}

\def\cf{{\it cf.~}}
\def\cc{CCs~}
\def\coll{ \tau_c}
\def\tma{ \tau_{ma}}
\def\tdr{ \tau_{dr}}
\def\tcr{ \tau_{cr}}
\def\tcs{ \tau_{cs}}
\def\ssp{c_{si}}
\def\roi{ \rho_i}
\def\ycm{ $Y_{cm}$~}
\def\pma{ \p_{max}}
\def\xcoor{{\it x}-coordinate~}
\def\ycoor{{\it y}-coordinate~}
\def\zcoor{{\it z}-coordinate~}
\def\ltsima{$\; \buildrel < \over \sim \;$}
\def\simlt{\lower.5ex\hbox{\ltsima}}
\def\gtsima{$\; \buildrel > \over \sim \;$}
\def\simgt{\lower.5ex\hbox{\gtsima}}

\title{3-D MHD Numerical Simulations of Cloud-Wind Interactions}

\author{G. Gregori\altaffilmark{1,2},
	Francesco Miniati\altaffilmark{2},
        Dongsu Ryu\altaffilmark{3},
and     T.W. Jones\altaffilmark{2}}

\altaffiltext{1}{Department of Mechanical Engineering, University of Minnesota,
    Minneapolis, MN 55455;
\\gregori@me.umn.edu}
\altaffiltext{2}{School of Physics and Astronomy, University of Minnesota,
    Minneapolis, MN 55455;
\\min@msi.umn.edu, twj@astro.spa.umn.edu}
\altaffiltext{3}{Department of Astronomy \& Space Science, Chungnam National
University, Daejeon 305-764, Korea;
\\ryu@canopus.chungnam.ac.kr}

\begin{abstract}

We present results from three-dimensional (3-D) numerical simulations investigating
the magnetohydrodynamics of cloud-wind interactions.
The initial cloud is spherical while the magnetic field is uniform and transverse
to the cloud motion.
A simplified analytical model that describes the magnetic energy evolution in front of
the cloud is developed and compared with simulation results.
In addition,
it is found the interaction of the cloud with a magnetized interstellar
medium (ISM) results in the formation of a highly structured magnetotail.
The magnetic flux in the wake of the cloud organizes into
flux ropes and a reconnection, current sheet is developed, as
field lines of opposite
polarity are brought close together near the symmetry axis.
At the same time, magnetic pressure is strongly enhanced at the
leading edge of the cloud from the stretching of the field
lines that occurs there. This has an important dynamical effect on the
subsequent evolution of the cloud, since some unstable modes tend to be strongly
enhanced.

\end{abstract}


\clearpage

\section{Introduction}
Magnetic fields are a pervasive element of the
interstellar and intergalactic medium and they are often
extremely relevant in characterizing local and global behaviors of these media.
Magnetohydrodynamic (MHD) cloud-wind interactions are believed
to be important in determing the observed filamentary and 
clumpy morphology associated
with clouds moving through a magnetized interstellar medium.
These processes may also result in a local enhancement of the
background magnetic field,
which, in turn, provides an important feedback on the 
subsequent evolution of the cloud. 
Jones \etal (1994, 1996) have shown that these regions of strong magnetic
pressure (what we call ``magnetic bumpers'') develop in front of the cloud as a result of
the stretching of the field lines that anchors on the cloud surface.
Such regions may also be adjacent to shocks, which can serve as acceleration sites of
high energy particles (\eg Jones \& Kang 1993). Within some supernova
remmants there is also clear evidence for supersonic clumps (\eg Jones
\etal 1998). Consequently the strong field regions can show enhanced
nonthermal radio emission (Jones \etal 1994). Miniati \etal (1997, 1999a)
investigated the role played by magnetic fields in cloud collisions by
comparing two dimensional (2-D) hydrodynamic and magnetohydrodynamic simulations. They
concluded that a magnetic field transverse to the cloud motion can
dramatically alter the outcome of the collisions, preventing the
disruption of the clouds otherwise occurring in almost all other
scenarios. Since the development of the magnetic
bumper was crucial for this result, Miniati \etal (1999b) have extended
the Jones \etal (1994, 1996) work. In particular, they further
investigated the formation of such a bumper under a broader range of
initial conditions by studying the propagation of
clouds through various oblique magnetic fields.
In addition they also assessed the issue of the
exchange of magnetic and kinetic energy during the evolution
of diffuse clouds in the ISM. There, in fact, they offer
a possible explanation for the comparable values of magnetic
and kinetic energy densities observed in some \ion{H}{1} complexes
\cite{obs1,obs2,obs3,obsm}.
Gregori \etal (1999) have presented three dimensional (3-D)
numerical simulations that show
dramatic dynamical effects of the magnetic field in determining the
cloud evolution during its propagation through a magnetized medium. In
particular, those authors found that a strong field enhances the development of
Rayleigh-Taylor unstable modes, thus hastening the cloud disruption.
On one hand this demonstrates that magnetic fields in 3-D
do not simply slip around the cloud surface, somewhat similar to the 2-D case.
On the other hand, the 3-D influence of the field was disruptive, opposite
to that seen in 2-D simulations.

In this paper we present 3-D numerical simulations
of moderately supersonic cloud motion in a magnetized interstellar
medium. The cloud is treated as non self-gravitating and adiabatic. First we 
present a detailed analytical model for the stretching mechanism of the field lines 
and the consequent magnetic field amplification at the cloud nose.
We also show that the final outcome in a cloud-wind
interaction is the development of complex features
analogous to the ones observed in cometary plasma tails, resulting
in the formation of a highly structured magnetotail. 
The magnetic
field itself organizes in coherent tails, or flux ropes, 
in the wake of the cloud. These ropes are associated with the development
of a Sweet-Parker reconnection sheet that alters the field topology
there. 
Our aim is to give a picture of the basic processes that
develop in cloud-wind interactions, in order to provide useful insight for
observations in a large variety of astrophysical
environments (see \eg Dgani \& Soker 1998).
The use of
3-D simulations is clearly a considerable advantage over previous work, since
now the full spatial domain can be investigated without the geometrical 
limitations
imposed by 1-D or 2-D calculations.  

The paper is organized as follows.
In \S\ref{code} we describe the numerical setup, and
the characteristic physical parameters of the problem.
The computational results are introduced in \S\ref{res}.
The development of the Rayleigh-Taylor instability
and the cloud disruption are briefly discussed in \S\ref{disr}.
The analysis of the magnetic energy evolution is given in
\S\ref{mag_evol}. In section
\S\ref{fr} we investigate the formation of flux ropes in the wake
of the cloud and in \S\ref{mrec} the magnetic reconnection.
Our results are summarized in \S\ref{cs}.

\section{Numerical Setup and Definition of the Problem}
\label{code}
The numerical computation is based on a total variation diminishing
(TVD) scheme for ideal MHD \cite{ryu1}. 
This is an explicit, conservative finite-difference
method with second order accuracy in space and time. 
We have used the multidimensional,
Cartesian version of the code (Ryu, Jones \& Frank 1995) with a constrained transport
scheme for preserving $\nabla \cdot {\bf B}=0$ \cite{ryu3}.
Neglecting self-gravity and radiative energy losses, the complete set of
the simulated equations for the
velocity ($\bf u$), magnetic field ($\bf B$), density ($\rho$), pressure
($p$) and ``color tracer'' ($C$) can be conveniently written as
\begin{equation}
   \frac{\partial \rho}{\partial t} + \nabla \cdot (\rho {\bf u}) = 0,
\end{equation}
\begin{equation}
   \frac{\partial {\bf u}}{\partial t} + {\bf u}\cdot\nabla {\bf u}
 	+\frac{1}{\rho} \nabla p -\frac{1}{\rho} (\nabla\times
	{\bf B})\times {\bf B} = 0,
\end{equation}
\begin{equation}
   \frac{\partial p}{\partial t} + {\bf u}\cdot \nabla p +
	\gamma p \nabla \cdot {\bf u} = 0,
\end{equation}
\begin{equation}
   \frac{\partial {\bf B}}{\partial t} - \nabla \times ({\bf u}
	\times {\bf B}) = 0.
\end{equation}
\begin{equation}
   \frac{\partial C}{\partial t} + {\bf u}\cdot \nabla C = 0.
\end{equation}
The last equation has been added to the standard set of ideal MHD
equations in order to be able to follow the motion
of the cloud material itself. $C$ corresponds, in fact, to the mass fraction
of the cloud gas inside the computational cell.

Initially, all the cloud material is labeled with
$C=1$, and the ambient medium with $C=0$. At any time $t$, 
the density of cloud
material in a fluid cell is given by $\rho_c = \rho\,C$, where $\rho$ is the total
fluid density at that point.  
In the MHD equations, the magnetic field is normalized such that the factor $4\pi$
does not appear, giving an Alfv\'en speed $v_A = B/\sqrt{\rho}$. We
assume
an adiabatic index $\gamma = 5/3$,
initial pressure equilibrium at $p_0 = 3/5$, and an initial density
in the background medium $\rho=\rho_i=1$.
Thus, the velocity is expressed in units of the sound speed in the
ambient medium: $c_s = (\gamma p_0/\rho_i)^{1/2}=1$.
The initial cloud density is $\rho_c=\chi \rho_i$, with $\chi=100$. A thin
transition layer $\sim 0.2 R_c$ ($R_c$ is the cloud radius) 
around the cloud, introduced to reduce the
Richtmyer-Meshkov instability at startup, 
brings the density to the value of the intercloud medium. 
A finite transition layer is in general expected due to both thermal 
condution (Balbus 1986) and photoionzation (Tielens \& Hollenbach 1985)
at the cloud boundary, although at this point
in our simulations it is considered
only for reasons of numerical stabililty.
The cloud is initially spherical in shape. 
The numerical value for its radius, $R_c$, is
set to unity, and this is chosen as the unit of length.
This also sets the unit time to the cloud sound crossing time, $\tau_{cs} = R_c/c_s$
($=1$ in numerical units).
At time $t=0$, the cloud is set in motion with respect to the uniform
background medium.
Its velocity is $u_c = M c_s$, with a starting value for the intercloud
Mach number, $M=1.5$.
The magnetic field is conveniently expressed in terms of the
familiar parameter
\begin{equation}
   \beta = \frac{p}{p_B},
\end{equation}
where $p_B=B^2/2$ is the magnetic pressure.
In our numerical simulations we have considered both the cases of an
initially strong field ($\beta=4$) and a weak field ($\beta=100$).
To be able to
compare with pure hydrodynamic effects, a case with $\beta=\infty$
(no magnetic field) has also been computed. 
A summary of all the simulations performed is given in Table \ref{tab1}.
In addition to the $\beta$ parameter, another important dimensionless
number often used to describe the action of the magnetic field on the
fluid motion is the Alfv\a'enic Mach number $M_A = u/v_{Ai}$, where
$v_{Ai} = B/\sqrt{\rho_i}$ is Alfv\a'en velocity in the intercloud
medium. So,
we initially have $M_A = 2.74$ for to $\beta=4$, and $M_A=13.7$
for $\beta=100$.

The computational domain is outlined in Fig. \ref{num_set}.
Symmetrical boundary conditions are employed
on the $y=0$ and $z=0$ planes, inflow conditions are applied on the
$x=0$ plane, while open conditions are used
on all other boundaries.
This choice eliminates odd modes of instabilities, but in companion, fully
3-D simulations at somewhat lower resolution we saw no evidence that such
modes play a deciding role in cloud evolution.
We employed a uniform grid,
$N_x \times N_y \times N_z = 416\times 208\times 416$, 
spanning $\slantfrac{1}{4}$ the volume of interest. The volume computed
was bounded along $x$-$y$, $x$-$z$ planes through the initial cloud
center. This gives a resolution of 26 zones per cloud radius,
less than that in our previous 2-D MHD
simulations (Miniati \etal 1999b, Jones \etal 1996).
So, small scale surface perturbations were relatively more damped.
However, from our experience, the adopted resolution is 
sufficient to capture basic cloud evolution over the time interval considered.
In addition, we have carried out several lower resolution 3D simulations
spanning the full volume of interest
(see Table \ref{tab1}). Some aspects of those companion simulations are 
reported as well in the following sections. The cloud shapes (sphere and cylinder)
that we have considered are quite ideal and they need to be considered to give only
a qualitative picture of the true interaction of a magnetized wind with
an interstellar cloud which has no distinctive shape. In this respect, the simulations
with spherical and cylindrical clouds show a quite similar behavior. Tests carried out
with a different shape, an elliptical cloud, have also confirmed the same pattern.
They all show qualitative agreement with the simulations reported here.

Since the cloud motion is supersonic, its motion leads to the formation of
a forward, bow
shock and a reverse, crushing shock propagating through the cloud.
The approximate time for
the latter to cross the cloud is referred to as the 
``crushing time''\footnote{
This form of the crushing time is a factor of 2 larger than the one of
Klein \etal
(1994), since our definition is based on the cloud
diameter instead of the cloud radius. We use this definition since it
more closely measures the actual time before the crushing shock emerges.}
(\eg Jones \etal 1994):
\begin{equation}
   \tau_{cr} = \frac{2 R_c \chi^{1/2}}{M c_s}.
\end{equation} 
Since the crushing time corresponds to the typical scale for the
cloud evolution, in the following figures and discussions, time will
be expressed in terms of $\tau_{cr}$.

Finally, it is necessary to add a few words about the significance of our
choice in the initial direction
of the field lines. In our simulations, the initial magnetic field has
been set up {\it transverse} (perpendicular) to the cloud motion. As
pointed out in previous 2-D simulations (Miniati \etal 1999b;
Jones \etal 1996; and Mac Low
\etal 1994), a magnetic field aligned with the direction of the cloud
motion never becomes directly dynamically relevant in terms of body
forces, even if it may have
some stabilizing
effects. For a general orientation, there will always be some component of the
field transverse to the cloud velocity, which will be stretched around
the cloud body. In this respect, our simulations
can be viewed as an approximate solution of the more general problem of
the oblique field orientation. 
As stressed by Jones \etal (1996) and Miniati \etal (1999b), 
for supersonic bullets most
field directions will indeed produce effects similar to the
transverse field case. 
To confirm that the field evolution in the experiments described here
are not special cases geometrically, we 
have also carried out fully 3-D simulations at low
resolution, including oblique initial field orientations, 
confirming the general behaviors for the more restricted symmetries
imposed.

\section{Results}
\label{res}

\subsection{Dynamical Evolution \& Cloud Disruption}
\label{disr}
As discussed in a companion paper (Gregori \etal 1999),
the magnetic field amplication that occurs in front of the cloud
inhibits instabilities in the $x$-$y$ plane,
but hastens those in the $x$-$z$ plane, thus accelerating the process
of cloud disruption.
An example of the effects of this 
{\it magnetically enhanced} Rayleigh-Taylor (R-T) instability
can be seen in Fig. \ref{sl_den}.
This shows density slices
in the $y=0$ plane for the $\beta=\infty$ (top), $\beta=100$ (middle) and
$\beta=4$ (bottom) simulations, at $t=0.94\tau_{cr}$ (left) and  $t=2.25\tau_{cr}$
(right).
Clearly, for the $\beta=4$ case, 
the growth of a R-T instability is much more rapid and, unlike the
other cases,
very pronounced density fingers have developed at the simulation end.
In addition, to illustrate how the cloud shape evolves,
we define generalized cloud sizes in terms of
the coordinate moments of inertia.
Following Klein \etal (1994) and Xu \& Stone (1995), 
the cloud extension in the $i$th direction, $R_i$, at time $t$ 
is given by
\begin{equation} \label{momin}
   R^2_i(t) = \frac{\left(\int r_i^2\,\rho\,C \,dV\right)_t}
   {\left(\int r_i^2\, \rho \, C\, dV\right)_0} R_c^2,
\end{equation} 
where $r_i$ is the $i$th position coordinate with respect to the center
of mass,
and the integrals are intended over the entire computational domain.
In Fig. \ref{pl_r}
we have plotted $R_y$ and $R_z$ for both the MHD and the
hydrodynamic simulations. Compared to the hydrodynamic case, the wrapping
of the field lines around the cloud in the $x$-$y$ plane produces 
a strong radial
magnetic pressure gradientgradient  that squeezes the cloud gas. Such pressure, consequently,
forces an extrusion of the cloud along the $z$ direction. This explains the
trends visible in Fig. \ref{pl_r}. There we can read that
for the $\beta=4$ case, after one
crushing time, $R_y$ has already been reduced by 30-40\%
while $R_z$ has been increased by a similar amount. 
We may observe that at 
$t \gtrsim 1.5 \tau_{cr}$, the expansion in the $z$ direction is sustained
at a very large rate, while $R_y$ stays almost constant.
Finally, in the $\beta=100$ simulation the change in the cloud form
follows the same qualitative pattern as for the $\beta=4$ case, 
although the evolution is considerably less dramatic and rapid.

\subsection{Magnetic Energy Evolution}
\label{mag_evol}
\subsubsection{Model}
In the attempt to understand the process of the
magnetic bumper formation and its interaction
with the surrounding flow, Miniati \etal (1999b) and
Jones \etal (1996) have proposed an approximate model
based on the Faraday induction equation.
Their analysis showed that the magnetic energy first increases with time
exponentially (Jones \etal 1996)
and then according to a power-law of index between one and two
(Miniati \etal 1999b).
In hopes that it may serve as a simple tool for understanding field growth
in such problems, we examine again the
magnetic energy evolution explicitly using
the flux freezing condition while incorporating the important
and more realistic assumption of a limited region of velocity
shear around the cloud nose.

In fact, the time evolution of the magnetic field, in ideal MHD 
is given by the solution of the following equation in Lagrangian
form
\begin{equation}
   \frac{d}{dt} \left(\frac{{\bf B}}{\rho}\right) =
   \frac{{\bf B}}{\rho} \cdot \nabla {\bf u}.
\end{equation}
This equation can be formally integrated as
(\eg Batchelor 1967) to give
\begin{equation}
    \left(\frac{{\bf B}}{\rho}\right) =
    \left(\frac{{\bf B}}{\rho}\right)_0 \cdot \frac{\partial}
    {\partial {\bf a}} {\bf X}({\bf a},t).
    \label{mag1}
\end{equation}
Here, $({\bf B}/\rho)_0$ refers to the value at time $t=0$, and
$\bf a$ is the initial position vector of the fluid element being followed. 
It is worth stressing that the previous equations are only valid far from shocks.
This is indeed the case in our analysis since
the magnetic energy tends to evolve mostly in front of the cloud, far from the bow shock and
the crushing shock penetrating into the cloud.
The mapping function of the flow field is
${\bf X}({\bf a},t)$, which locates the 
fluid element at subsequent times.
Its derivative, $\partial {\bf X}/\partial {\bf a}$,
parametrically measures
the deformation of a fluid element, and, therefore {\it determines the local
growth of the magnetic field}. From eq. (\ref{mag1})
the strength of the magnetic field, $|{\bf B}|=B$, 
then evolves according to: 
\begin{equation}
    \left(\frac{ B}{\rho}\right) =
    \left(\frac{{B_i}}{\rho}\right)_0 \frac{\partial}
    {\partial {a_i}}  X({\bf a},t)
     \simeq \left(\frac{B}{\rho}\right)_0
    \frac{\partial X} {\partial a},
    \label{mag2}
\end{equation}
where $X=|{\bf X}|$ and $a=|{\bf a}|$.
In order to derive the evolution of the term $\partial X/\partial a$
we assume steady flow and 
write the equation for a fluid element along a flow line as
(\eg Aris 1962): 
\begin{equation}
    \frac{ d(dX)}{dt} = \frac{\partial u_i}{\partial x_j}~ \frac{ dx_j}{dX} 
~\frac{ dx_i}{dX} ~dX \simeq
    \left(\frac{{\partial u_x}}{\partial y}+\frac{{\partial u_y}}{\partial
x}\right) \cos\theta \sin\theta ~dX,
\label{dxeq}
\end{equation}
where we neglect the terms involving $u_z$ and indentify $dx_j/dX$ and 
$dx_i/dX$ as the direction cosines of the flow line. 
To integrate equation (\ref{dxeq})
we make the important and realistic assumption that the velocity shear
vanishes outside a finite region in front of the cloud. 
This assumption is confirmed by Fig.
\ref{vel_hb} where we can see that in most cases the velocity field is
characterized by a strong shear only around the cloud front. 
We emphasize that 
{\it if a fixed portion of a fluid line is
subject to stretching, then its length, and, hence, its strength, 
will grow only linearly} (as shown below by eq.
\ref{dxeq1}), {\it as opposed to an exponential increase
occuring when the stretching is
constant over the full extent of the fluid line.}
In addition, we
assume for simplicity that such a shear pattern is approximately constant 
in space. Therefore, 
integrating eq. (\ref{dxeq}) in both space and time we obtain in that
restricted region:
\begin{equation}
    X({\bf a},t) \simeq \int_0^t dt \int_a^{a+\ell _a}
    \left(\frac{{\partial u_x}}{\partial y}+\frac{{\partial u_y}}{\partial
x}\right) \cos\theta \sin\theta ~dX \simeq a+\frac{1}{2} \ell _a
\left(\frac{{\partial u_x}}{\partial y}+\frac{{\partial u_y}}{\partial
x}\right) t
\label{dxeq1}
\end{equation}
where $\ell _{a}$ represents the length of the flow line over which the
velocity shear is non-null and the term $\sin\theta
\cos \theta$ has been taken to contribute a factor \slantfrac{1}{2} coming
from integrating along an arc size of the order of
\slantfrac{\pi}{2}, as appropriate for the cloud nose.
We are now able to derive the quantity of interest, namely $\partial X/\partial a$.
Assuming a laminar flow regime before the cloud and around its nose, 
it is possible to see that the change of the length (along which
the strain occurs) $\delta \ell _a$ between two adjacent flow
lines is of the order of their separation $\delta a$, or, equivalently
$\delta \ell _a / \delta a = 1$.
We approximate also the velocity shear
\slantfrac{1}{2} $(\partial v_x/\partial y+\partial v_y/\partial x)\sim
\lambda(t) u_c/R_c$,
since the flow speed increases from zero on the cloud nose
to roughly the cloud speed, $u_c$,
along an arc of length $\sim R_c$ at the start of the simulation.
Here, $\lambda(t)$ is a slowly varying quantity introduced to correct
for all the details of the MHD flow dynamics that have not been included in this
simplyfied analysis.
From equations (\ref{mag2}) and (\ref{dxeq1}) we have then
\begin{equation}
   \frac{ B}{\rho} \simeq \frac{B_0}{\rho_0} \left(1 + \lambda 
\frac{u_c}{R_c} t
   \right),
\end{equation}
or in terms of the magnetic energy density (magnetic pressure)
$p_B = B^2/2$,
\begin{equation}
   p_B \simeq p_{B 0} \left(\frac{\rho}{\rho_0} \right)^2 
   \left(1 + \lambda \frac{u_c}{R_c} t \right)^2.
   \label{mag3}
\end{equation}
The quantity $\lambda(t)$ parameterizes the evolution in the
flow. In the beginning this mostly represents an expansion of the
flow field in response to increased magnetic pressure; that is,
the flow lines become more spread out. Eventually,
however, as the cloud begins to decelerate, it includes reductions in
its asymptotic speed, as well.
The details of the cloud dynamics embodied in $\lambda(t)$ are difficult
to estimate. Nevertheless, based on the results of our 3-D simulations and on the
theoretical model of Miniati \etal (1999b), we believe that for
$t > \tau_s=R_c/u_c$ a functional
form $\lambda(t) \sim (t/\tau_s)^{-q}$ (with $q\sim \slantfrac{1}{2}$)
is adequate to capture the main features in the magnetic field evolution.
Then, we can rewrite
eq. (\ref{mag3}) as becoming asymptotically
\begin{equation}
   p_B \simeq p_{B 0} \left(\frac{\rho}{\rho_0}\right)^2
   \left(\frac{t}{\tau_s}\right)^{2 (1-q)}.
   \label{mag4}
\end{equation}
For $q=\slantfrac{1}{2}$ the magnetic pressure would continue to grow linearly
with time. As discussed below, our simulations indeed suggest that the field
pressure tends to increase with a power law exponent of order unity, at least until the
crushing shock has completely crossed the cloud.

\subsubsection{Quantitative Analysis and Comparison with Numerical Simulations}
Neglecting the initial times ($t \lesssim \tau_s$), when
compressibility is dominant, in the following we will assume
$\rho \approx \rho_0$ and only consider the time 
regime in which equation 
(\ref{mag4}) is appropriate in the flow around the cloud.
Then, let us estimate the integral form of equation (\ref{mag4}) over the
entire computational volume $V$.
We indicate with $V_{\epsilon}$ the region around the cloud where the
velocity shear is large, and, therefore, field amplification takes place.
Then $\lambda u_c/R_c$ becomes the average rate of strain over
such a volume, outside of which the magnetic field remains approximately
equal to its initial value. Typically, $V_{\epsilon}$ is of the order of
$V_c$, the cloud volume.
\begin{equation}
   E_B = \int_V p_B dV \simeq
      p_{B0} V_c \left(\frac{t}{\tau_s}\right)^{2 (1-q)} +
      p_{B0} (V-V_c).
   \label{m_mag1}
\end{equation}
Then, we can rearrange (\ref{m_mag1}) as 
\begin{equation}
   \frac{E_B - E_{B0}}{E^{c}_{B0}} \equiv \frac{\Delta E_B}{E^{c}_{B0}}
   \simeq 
   \left(\frac{t}{\tau_s}\right)^m,
\end{equation}
where $m=2 (1-q)$,
$E_{B0} = p_{B0} V$ is the initial magnetic energy in the
computational volume and 
$E^c_{B0} = p_{B0} V_c$ is the initial cloud magnetic energy. 
In Fig. \ref{pl_eb}  (top panel) we plot the time dependence of
$\Delta E_{B}/E^c_{B0}$ obtained directly from the numerical simulation
to compare with this relation.
A lower resolution simulation result
(16 zones per cloud radius) is also given for the $\beta=4$ case. Both
the high and the low resolution curves are sufficiently close to suggest
a good convergence for the global magnetic energy evolution.   
It is clear that the total magnetic energy in the volume tends to
increase monotonically over the simulated time.

Fitting the
numerical results in the time
interval $0.01 \lesssim t/\tau_{cr} \lesssim 1.0$ to a $t^m$ dependence,
gives a power law
index $m \approx 1.1$ for $\beta=4$ and 
$m \approx 1.4$ for $\beta=100$.
These results suggest $m \rightarrow 1$, at least during
this interval in the cloud evolution.
Indeed, a linear increase in the magnetic energy is to be
expected.
In fact, up to $t \simeq \tau_{cr}$ the cloud does not suffer any deceleration.
Since the system is then supplied with a steady flux of kinetic energy, we may expect
that a constant fraction of the latter is converted into magnetic energy
at a constant rate. 
However, Fig. \ref{pl_eb} shows that, in both cases, at $t \gtrsim \tau_{cr}$
the rate of magnetic energy growth drops below linear. This is a signature
of the beginning of the cloud deceleration and the consequent reduction in
energy supply, although some outflow of magnetic energy from the computational
volume is also occuring (see below).
It is of interest to compare the increment in magnetic energy to the
evolution of the 
total kinetic energy inside the computational box. This is also shown
in Fig. \ref{pl_eb} (bottom panel). From it we learn that at about
one crushing time more than 10\% of the initial kinetic energy has been converted into
magnetic energy for the $\beta=4$ case. This fraction becomes larger that 15\%
at simulation end. On the other hand, for the $\beta=100$ case, the conversion
amounts to only a few percent, in accordance with a much larger
conversion timescale as determined later (see \S \ref{timescales}).
We point out that the increments in the magnetic energy correspond to 
and are responsibile
for the decrements over time of the kinetic energy
(measured with respect to an observer who sees the intercloud medium at
rest) for both values of $\beta$ as plotted in Fig. \ref{pl_ek}.
In reality, a closer analysis reveals that at simulation end, the decrements
in kinetic energy are somewhat larger than the
correspondent magnetic energy increments, indicating that some magnetic
energy flux has escaped the computational box.
Thus, the final values for the $\Delta E_B$ in Fig. \ref{pl_eb} are
understood only as lower limits.
Comparing results at different resolutions for
the $\beta=4$ case, again indicates a good convergence of the simulations
in this property.
It is worth mentioning that this strong increase in the magnetic pressure in
front of the cloud turns out to be a result rather independent of the
initial cloud geometry. In fact, Maxwell stresses will tend to squeezed
the cloud into a cylinder-like structure with
the axis aligned perpedicular to the plane defined by the
initial field and the direction of motion. In this respect, low resolution
simulations started with a cylinder as the initial cloud shape, revealed
a qualitatively similar behavior in the magnetic energy evolution, although an
even stronger increase in the magnetic pressure occured at the cloud nose.
Overall these results compare well with the previous 2-D
simulations of Miniati \etal (1999b). While the increase of magnetic
energy in the 3-D simulation is slightly lower than for the 2-D case,
the values are comparable in order of magnitude.

\subsubsection{Timescales}
\label{timescales}
In the evolution of the cloud several dynamical timescales
can be identified. As the magnetic pressure in front of the
cloud increases, the flow progressively tends to be dominated
by Maxwell stresses and its global behavior radically changes
from the pure hydrodynamical case. Typically, this occurs
when the magnetic pressure becomes comparable to the ram pressure of
the gas. Gregori \etal (1999) have shown that the onset of this
transition is characterized by magnetically enhanced R-T modes that
ultimatively disrupt the entire cloud. 
In Fig. \ref{pl_pb} the ratio of the {\it maximum} magnetic pressure, $p_B$,
along the symmetry axis ($y=z=0$)
with respect
to the ram pressure of the intercloud fluid, $\rho_i u_c^2$, is plotted 
as a function of time for the two values of $\beta$.
At the beginning of the simulation $p_{B0}/(\rho_i u_c^2)
\approx 0.067$ for $\beta=4$ and $p_{B0}/(\rho_i u_c^2)
\approx 0.003$ for $\beta=100$. Thus, the ram pressure is initially
dominant. However, as shown in Fig. \ref{pl_pb} for the
$\beta=4$ simulation, at about one crushing time
the magnetic pressure maximum becomes comparable and {\it even larger than} the
initial ram pressure. Its subsequent decrease is connected to
the rapid development of the R-T instability;
in fact the field lines move away from the symmetry axis, towards the
newly developing indentations (see Fig. \ref{sl_den}). Our analysis 
of the field inside such structures reveals that the 
magnetic pressure keeps growing there 
and at $t \sim 2 \tau_{cr}$
reaches a ratio of 1.66 with respect to the ram pressure.
For the $\beta=100$ case we note that, despite a
remarkable increase with respect to its initial 
value, the magnetic pressure maximum remains well below the 
ram pressure limit throughout the simulation time. 
Based on the model developed in the previous sections, 
after the initial transient, from equation (\ref{mag4}) we estimate
\begin{equation}
   \frac{p_B}{\rho_i u_c^2} \simeq \frac{2}{M_A^2} \left(
   \frac{t}{\tau_s}\right)^m,
\end{equation}
where $M_A$ is the initial Alf\a'enic Mach number. Using $m \sim 1$,
we estimate that the magnetic pressure becomes comparable to the initial ram pressure
for $t \gtrsim (M_A^2/2)^{\frac{1}{m}} \tau_s \sim
(M_A^2/2) \tau_s$, or $t \gtrsim 0.28 \tau_{cr}$ ($\beta=4$) and
$t \gtrsim 7.03 \tau_{cr}$ ($\beta=100$). This is indeed confirmed from Fig.
\ref{pl_pb}. For the $\beta=4$ simulation at $t>0.3 \tau_{cr}$ the
magnetic pressure maximum is already comparable with the ram pressure.
On the other hand, in the weak field
case, the magnetic pressure maximun remains much below the initial ram pressure even at
simulation end. These results roughly confirm our
findings for the behavior of the magnetic field at the cloud nose.

Another significant timescale can be defined
in terms of the energy exchange between kinetic and magnetic energy,
since the stretching of the field lines acts
as a conversion mechanism in which the cloud kinetic energy is transformed into
magnetic energy.
By analogy with Miniati \etal (1999b) we can estimate the time
required for the magnetic energy to
equal the initial kinetic energy of the cloud by solving the equation
\begin{equation}
	\frac{1}{2} \rho_c u_c^2 V_c = E_B-E_{B0} \simeq p_{B0} V_c \left(
	\frac{t}{\tau_s}\right)^m.
\end{equation}
We easily get the characteristic
timescale
\begin{equation}
	\tau_{ma} = \left(\frac{\chi M_A}{4}\right)^{1/m} \tau_s
	\simeq \left(\frac{\chi M_A}{4}\right) \tau_s.
	\label{t1}
\end{equation}
where the last term has been obtained by setting $m=1$, as
previously observed. In this respect, $\tau_{ma}$ must be intended
as a characteristic {\it braking time}, since it sets an upper limit
for the cloud motion across the ISM.
We should note that
in absence of magnetic field, the characteristic time required to stop the
cloud is given by $\tau_d = \chi \tau_s$ (\eg Jones \etal 1994).
It is then evident that the cloud motion is magnetically dominated only if
$\tau_{ma} \lesssim \tau_d$. By comparison with (\ref{t1}), we see that this
is the case if $M_A < 4$. This indicates that
for $\beta=4$ ($M_A = 2.74$) the magnetic field plays a dynamically important
role from the beginning.
On the other hand, in the $\beta=100$ case, $M_A = 13.69$ and 
the hydrodynamic drag is the relevant timescale in determining the cloud
evolution; thus in this case magnetic effects are likely not to
be important.
Mouschovias \& Paleologou (1979) and Elmegreen (1981) have calculated the
characteristic time required by the magnetic field to stop the cloud
using a different approach based on the radiated energy through
Alfv\a'en waves.
Typically, for transverse field geometry 
and an infinitely long cylindrical cloud, they estimate that the
timescale for magnetic braking is of the order of 
$\slantfrac{1}{2}\, \chi M_A \tau_s$ which, apart from
a numerical factor, 2, is equivalent to our result.

The timescale (\ref{t1}) can be recast in a dimensional form:
\begin{equation}
   \tau_{ma} = 7.04\times 10^{10} \chi \beta^{1/2}
   \left(\frac{R_c}{0.1\,{\rm pc}}\right)
   \left(\frac{10^6\,\rm{cm/s}}{c_s}\right)\,{\rm s}.
\end{equation}
The characteristic time for radiative cooling is given by (Spitzer 1978)
\begin{equation}
   \tau_{cool} = \frac{3}{2}\frac{k_B T}{n_i \Lambda} \simeq 
   1.6\times 10^{14}\,{\rm s},
\end{equation}
where $n_i$ is the number density in the intercloud medium and
$\Lambda$ is the interstellar cooling function. For temperatures of the interstellar gas in
the order of $10^4$ K and densities $n_i \sim 0.1\,{\rm cm}^{-3}$, we have
$n_i \Lambda \sim 1.3\times 10^{-26}$ erg/s (Ferrara \& Field 1994) to give
the numerical value in equation (3-16).
We clearly see that for the values of the parameters
$\chi$ and $\beta$ used in the simulations, the evolution of
the system remains always strictly adiabatic (that is, $\tau_{ma} < \tau_{cool}$) only for
for clouds with radius $R_c \lesssim 0.1$ pc. 
In the case of larger clouds, radiative effects have still a small influence
on the cloud dynamics
if the cooling time is larger than the characteristic time for
re-expansion of the postshock interstellar material as it flows around
the cloud. Such time is in the order of $R_c/c_s$ (Miniati \etal 1997).
From the condition $\tau_{ma}>R_c/c_s$ we then obtain that cooling
remains negligible for clouds radii $R_c \lesssim 10$ pc.

\subsection{Flux Ropes}
\label{fr}
When the magnetized fluid flows into the cloud wake, the field lines frozen
in it produce elongated structures of strong
field concentration surrounded by a thin vortex and
current sheet. These
filamentary structures are clearly visible in Figs. \ref{sl_all_b100}
and \ref{sl_all_b4}.
Following Mac Low \etal (1994) we identify these regions 
as ``flux-ropes'', where individual flux tubes are organized in a
coherent pattern. 
In the simulations presented here,
however, we see no signs of twist in them.
The complete 3-D structure is 
illustrated in Fig. \ref{den_rope} for $t = 0.94\tau_{cr}$. 
In both the the strong and the weak
field simulations, the flux ropes converge toward the
symmetry axis where the thermal pressure is low (Mac Low \etal 1994). 
There are then strong similarities
between our results and the 2-D simulations of Mac Low \etal (1994)
and Jones \etal (1996), which show the same flux rope structures
for the transverse field geometry.
From Fig. \ref{sl_all_b100} we can see that in the weak field
($\beta=100$)
simulation the flux rope remains very close to the symmetry
axis for almost its entire length. Inside the flux rope the
field strength at this time ($t=0.94 \tau_{cr}$)
is comparable to that at the leading edge of the cloud. In the cloud wake the dominant
enhancement of the field strength is probably produced by the compression
of the rope on the symmetry axis.
Such compression, in turn, is due to the large gradient of thermal pressure
produced there by the cloud motion.
We should also note that some additional field amplification
may occur as the fluid crosses the tail shocks.
Conversely, in the strong field, $\beta=4$, simulation (Fig. \ref{sl_all_b4}), 
the flux rope is kept close to
the symmetry axis only in a small region just at the cloud rear
(a few cloud diameters in size). Farther away it opens up in ``wings''
that tend asymptotically to align with the initial unperturbed 
field direction. Finally, the magnetic field value (at $t=0.94\tau_{cr}$,
$\beta=4$) in these ropes
is on average a factor $\sim 2$ larger than the background field.

As we can see in Fig. \ref{vel_hb}, an important feature of the
flux ropes is that the plasma in them is dynamically tied to the cloud; that is, the
plasma there moves with the cloud.
This is a very
remarkable characteristic because the flux ropes, formed by
magnetic field lines stretched around the shape of the cloud, do not
carry cloud material
{\it per se} (demonstrated through the tracer variable $C$).
The separation of the flux ropes from the external flow occurs
through a tangential discontinuity, with the magnetic field parallel
to the interface and no mass flow across it. Such a discontinuity is
characterized by a constant total pressure (magnetic plus
thermal) across the separation interface. However, the higher magnetic
pressure inside the rope is balanced by a considerably lower density
there, compared to the outside fluid material.
As already noted, there is little or no twist generated by magnetic
or current helicity in the flux ropes. They are sheet-like. Similar
sheet-like morphologies were noted in strong magnetic structures formed
in high resolution 3-D MHD Kelvin-Helmholtz instability simulations (Ryu \etal 2000).
In that case it was clear that the sheet-like morphology was a consequence
of the strong magnetic field and not a resolution effect. This may be true here as
well, but our resolution is not sufficient to assess that.

\subsection{Magnetic Reconnection}
\label{mrec}
The ideal MHD approximation breaks down in regions where resistive effects
or diffusion dominate the field evolution. Even if the scale length of these
regions is very small, their development may change the global topology
of the field lines (Song \& Lysak 2000). 
The most important of such events is magnetic
reconnection. This has been observed in the heliosphere and in various
2-D and 3-D numerical simulations
(\eg Biskamp 1994, Miniati \etal 1999b, Antiochos \&
De Vore 1999, Ryu \etal 2000). 
Miniati \etal (1999b), using the 2-D version
of the MHD code applied here,
showed clear 2-D reconnection in the wake of a cloud, apparently
due to tearing mode instabilities. 
Our new numerical
simulations confirmed the previous results for clouds and showed that, indeed,
reconnection takes place in the wake of the cloud. In fact from Figs.
\ref{sl_all_b100} and \ref{sl_all_b4}, it is clear
that on the symmetry plane, 
$y=0$, a pair of thin current sheets form and that 
field annihilation occurs there. 
Here, the reconnection process seems steady, however, that can be understood by examining
the properties of the reconnection region.
The two figures show that the ratio 
$L/\delta$ between the length and the full thickness of the current sheet is
$\sim 15$-$25$ for both the strong and weak field simulations.
The current sheet structure, for the $\beta=4$ case, reveals strong
similarities with the classic 2-D Sweet-Parker current sheet model
(\eg Biskamp 1993). In this respect, the recent 2-D 
resistive MHD simulations of Uzdensky \& Kulsrud (1999) also seem to
confirm that reconnection tends spontaneously to evolve toward a 
Sweet-Parker steady state. Additional analogies with other reconnection models,
\eg Syrovatskii's solutions (Biskamp 1993), can also
be found in the branching of the ``separatrix'' on the right side of the
sheet.
In the presence of a reconnection layer the issue
regarding its stability immediately is raised. Stability against
tearing modes is expected if
$S \sim (L/\delta)^2 \lesssim 10^4$ (Biskamp 1993), where $S$ is the
Lundquist number of the current sheet. This is indeed in agreement
with our present simulations, which show no development of such an
instability.
That contrasts as expected with our previous 2-D simulations where 
the
tearing mode instability was, in fact, observed in the analogous regions
Miniati \etal (1999b). In order to understand the difference between the two
cases, we recall that the Lundquist number is actually defined as
\begin{equation}
   S = \frac{v_A L}{\eta},
\end{equation}
where $v_A$ is the Alfv\a'en velocity 
of the magnetic field on the top and bottom of the current sheet and
$\eta$ is the fluid resistivity. 
In the Sweet-Parker reconnection model $S$ takes the form $S=(L/\delta)^2$. 
In simulations of ideal MHD $\delta$ is
limited by numerical discretization and its value is $\sim 2$-$3 \Delta x$,
where $\Delta x$ is the size of a numerical zone.
The length of the reconnection region in the cloud wake will generally
be several cloud radii, typically 5-10. 
This implies that the values for $L/\delta$ are numerically
limited: $S \lesssim (2$-$3\times N_c)^2$, where $N_c$ is the number of
numerical zones per cloud radius.
In our current 3-D simulations, $N_c = 26$, so 
$S < 10^4$, and the layer should be stable to tearing modes as, indeed, it is.
On the other hand, our 2-D simulations were done at roughly twice the
resolution across the clouds, with $N_c = 50$, so the current sheet
in the tail was characterized by $S > 10^4$. It should have been,
and was, tearing mode unstable.

\section{Summary \& Concluding Remarks}
\label{cs}
In this paper we have presented the results of a series
of 3-D MHD numerical simulations of cloud-wind interactions.
We have considered an initial spherical cloud
that moves transverse to the magnetic field, with two different
cases for its initial strength; namely $\beta = p/p_{B} = 4$ and $\beta = 100$.
Both the weak ($\beta = 100$) and strong field ($\beta = 4$) 
simulations showed a qualitatively comparable behavior with
a substantial enhancement of the magnetic pressure at the leading edge
of the cloud.
This confirms and extends previous 2-D results
(Jones \etal 1996, and Miniati \etal 1999b).
A new, detailed analysis of the field stretching that occurs in front of the
cloud is presented here. Qualitative agreement
with the numerical simulations is obtained.
In particular, it is shown that magnetic
effects tend to be important only if the initial Alfv\a'enic Mach number
is sufficiently small. In this case the cloud dynamical
evolution (acceleration and disruption) occurs on shorter timescales and
a substantial rapid conversion of kinetic into magnetic energy takes place.
Thus, at $t \lesssim \tau_{cr}$, the magnetic pressure becomes comparable with
or greater than the ram pressure.
On the other hand, if the initial Alfv\a'enic Mach number is large, then the
evolution of the cloud is still dominated by the hydrodynamic drag and
conversion of kinetic into magnetic energy occurs at a much smaller rate.

For typical values of the magnetic field in the ISM, 
the propagation of a cloud through a magnetized medium should produce
inhomogeneities
on a scale comparable to the cloud size. As already pointed out in Miniati
\etal (1999b), this
could help explain observed fluctuations in the
galactic magnetic field (Heiles 1989, Meyers \etal 1995).
Observations of Gloeckler \etal (1997)
in the local interstellar cloud also seem to suggest the possibility
of an inhomogeneous distribution of the field intensity.
Moreover, we illustrate a complex series of
topological modifications that take place in the cloud wake and 
characterize the dynamics of the flow there.
In the tail of the cloud the field lines aggregate in a coherent
pattern to form long, sheet-like flux ropes. The plasma entrained there
moves approximately with the cloud. Additionally, 
along the symmetry axis, magnetic reconnection takes place.
Evolutionary details such as 
the formation of vortical structures at the cloud interface, 
as well as the development of instabilities of the
reconnection current sheet were prevented by the limited
numerical resolution.
Despite these limitations, the simulations reported here
give a significant
picture of the fundamental processes governing cloud-wind
interaction in a magnetized ISM, 
setting important constraints for further
observational and theoretical studies. In this respect, our results
may give support to recent models of cloud-wind interactions such as 
those proposed to explain the
formation of nonthermal filaments in the galactic center
(Shore \& LaRosa 1999). Investigations 
of planetary nebulae interacting with a magnetized fluid may also
take advantage of the results presented here (see \eg
Soker \& Dgani 1997, Dgani \& Soker 1998).

\acknowledgments
This work is supported at the University of Minnesota by the NSF through
grants AST 96-19438 and INT95-11654, by NASA grant NAG5-5055, and by the
Minnesota Supercomputing
Institute. Work by D.R. is supported in part by Korea Research
Foundation Grant KRF-99-015-DI0113. We thank the referee, Dinshaw 
Balsara, for helpful comments on improving the manuscript.

\clearpage

\begin{center}
{\bf FIGURE CAPTIONS}
\end{center}

\figcaption[]{Schematic of the numerical setup.
\label{num_set}}


\figcaption[]{Slice of $log(\rho)$ on the plane $y=0$ at $t=0.94\tau_{cr}$ 
(left column) and $t=2.25 \tau_{cr}$ (right column). Top: $\beta=\infty$,
center:
$\beta=100$, bottom: $\beta=4$.
\label{sl_den}}

\figcaption[]{Relative cloud momenta in the $y$ and $z$ directions (see
text).
\label{pl_r}}

\figcaption[]{Top: magnetic energy increment in the computational volume normalized with
respect to the initial magnetic energy inside the cloud ($\Delta E_B/E^c_{B0}$). 
Bottom: magnetic energy increment in the computational volume normalized with
respect to the initial kinetic energy of the cloud ($\Delta E_B/E^c_{K0}$).
The high resolution (hr) simulation corresponds to
case 3 in Table 1 and the low resolution (lr) simulation corresponds
to case 4 in Table 1.
\label{pl_eb}}

\figcaption[]{Total kinetic energy in the computational volume normalized with
respect to its initial value ($E_K/E_{K0}$). The high resolution (hr) simulation
corresponds to case 3 in Table 1 and the low resolution (lr) simulation
corresponds to case 4 in Table 1.
\label{pl_ek}}

\figcaption[]{Maximum value of the magnetic pressure on the symmetry
axis ($y=z=0$) with respect to its initial ram pressure of the fluid.
\label{pl_pb}}

\figcaption[]{Slice in the plane $z=0$ for the $\beta=100$ simulation at
$t=0.94 \tau_{cr}$. 
$\log(\rho)$
top left, $\log(B^2/2)$ top right, $\log(\omega^2+1)$ bottom left,
$\log(j^2+1)$ bottom right. Here, ${\bf \omega}=\nabla\times{\bf u}$ 
is the vorticity and ${\bf j}=\nabla\times{\bf B}$ is the current density.
The $B_x$ and $B_y$ components of the magnetic field are represented by the
arrows.
\label{sl_all_b100}}

\figcaption[]{Slice in the plane $z=0$ for the $\beta=4$ simulation at 
$t=0.94 \tau_{cr}$.
$\log(\rho)$
top left, $\log(B^2/2)$ top right, $\log(\omega^2+1)$ bottom left,
$\log(j^2+1)$ bottom right. Here, ${\bf \omega}=\nabla\times{\bf u}$ 
is the vorticity and ${\bf j}=\nabla\times{\bf B}$ is the current density.
The $B_x$ and $B_y$ components of the magnetic field are represented by the
arrows.
\label{sl_all_b4}}

\figcaption[]{Left column: volume rendering of the magnetic pressure (log scale)
for the $\beta=100$ simulation (top) and $\beta=4$ (bottom).
Right column: volume rendering of the cloud density (log scale) 
for the $\beta=100$ simulation (top) and $\beta=4$ (bottom).
Time is expressed in units of $\tau_{cr}$.
\label{den_rope}}

\figcaption[]{Slice of $\log(\rho)$ on the plane $z=0$ at $t=0.94 \tau_{cr}$ (left
column) and $t=2.25 \tau_{cr}$ (right column). Top: $\beta=\infty$, center:
$\beta=100$, bottom: $\beta=4$. The $u_x$ and $u_y$ components of the
velocity field are represented by the arrows.
\label{vel_hb}} 

\clearpage

\begin{deluxetable}{ccccccccc}
\footnotesize
\tablecaption{Summary of 3-D MHD cloud simulations.
\label{tab1}}
\tablewidth{0pt}
\tablehead{
\colhead{Case} & \colhead{Symmetry} &
\colhead{Resolution\tablenotemark{{\rm b}}} &
\colhead{Initial field} &
\colhead{Shape} & \colhead{$M$} & 
\colhead{$\chi$} & \colhead{$\beta$} &
\colhead{$M_A$}  \\
&  \colhead{plane(s)\tablenotemark{{\rm a}}} & & \colhead{direction} & \colhead{($l_{x},l_{y},l_{z}$)
\tablenotemark{{\rm c}}} &
 &  &  &  
}
\startdata
1  & $x$-$y$, $y$-$z$  & 26  & (0,1,0)  & (1,1,1)  & 1.5 &  100 & $\infty$ &  $\infty$  \\
2  & $x$-$y$, $y$-$z$  & 26  & (0,1,0)  & (1,1,1)  & 1.5 &  100 & 100      &  13.7      \\
3  & $x$-$y$, $y$-$z$  & 26  & (0,1,0)  & (1,1,1)  & 1.5 &  100 & 4        &  2.74      \\
4  & $x$-$y$, $y$-$z$  & 16  & (0,1,0)  & (1,1,1)  & 1.5 &  100 & 4        &  2.74      \\
5  & none	       & 16  & ($\frac{1}{\sqrt{3}}$,$\frac{1}{\sqrt{3}}$,$\frac{1}{\sqrt{3}}$)
                                        & (1,1,2)  & 3.0 &  100 & 10       &  8.66      \\
6  & none              & 16  & ($\frac{1}{\sqrt{3}}$,$\frac{1}{\sqrt{3}}$,$\frac{1}{\sqrt{3}}$)
                                        & (1,1,2)  & 3.0 &   10 & 10       &  8.66      \\
7  & none              & 16  & ($\frac{1}{\sqrt{3}}$,$\frac{1}{\sqrt{3}}$,$\frac{1}{\sqrt{3}}$)
		                        & (1,1,2)  & 3.0 &   30 & 10       &  8.66      \\
8  & none	       & 16  & ($\frac{1}{\sqrt{3}}$,$\frac{1}{\sqrt{3}}$,$\frac{1}{\sqrt{3}}$)
		                        & (1,1,2)  & 3.0 &  100 &  2       &  3.87      \\
9 & none              & 16  & ($\frac{1}{\sqrt{3}}$,$\frac{1}{\sqrt{3}}$,$\frac{1}{\sqrt{3}}$)
		                        & (1,1,2)  & 3.0 &  100 &  4       &  5.48      \\		                        		                         	                       
	
\enddata


\tablenotetext{a}{plane(s) across which symmetric boundary conditions are
employed in the numerical computations.}
\tablenotetext{b}{in number of zones per cloud radius, $R_c$.}
\tablenotetext{c}{initial cloud axis in units of $R_c$. (1,1,1) corresponds
to a sphere and (1,1,2) corresponds to a cylinder with the axis in the
$z$ direction.}

\end{deluxetable}

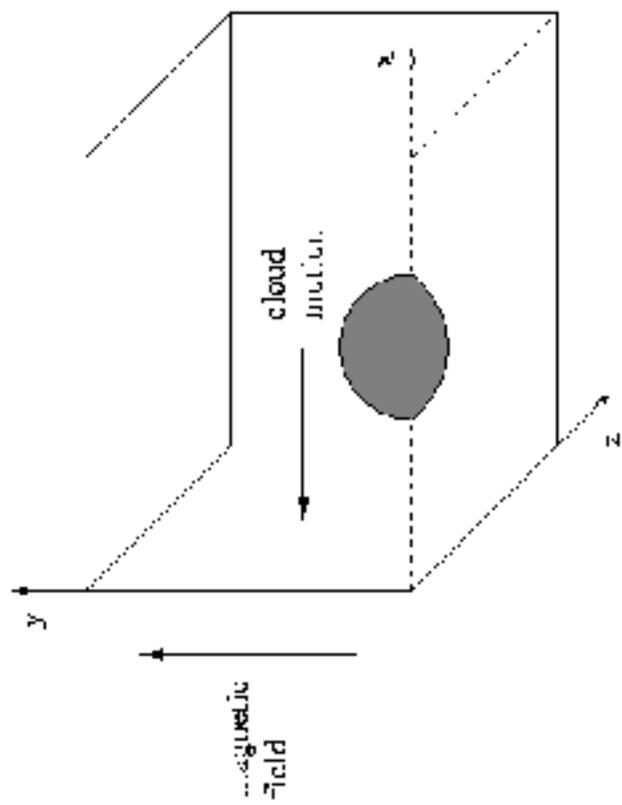
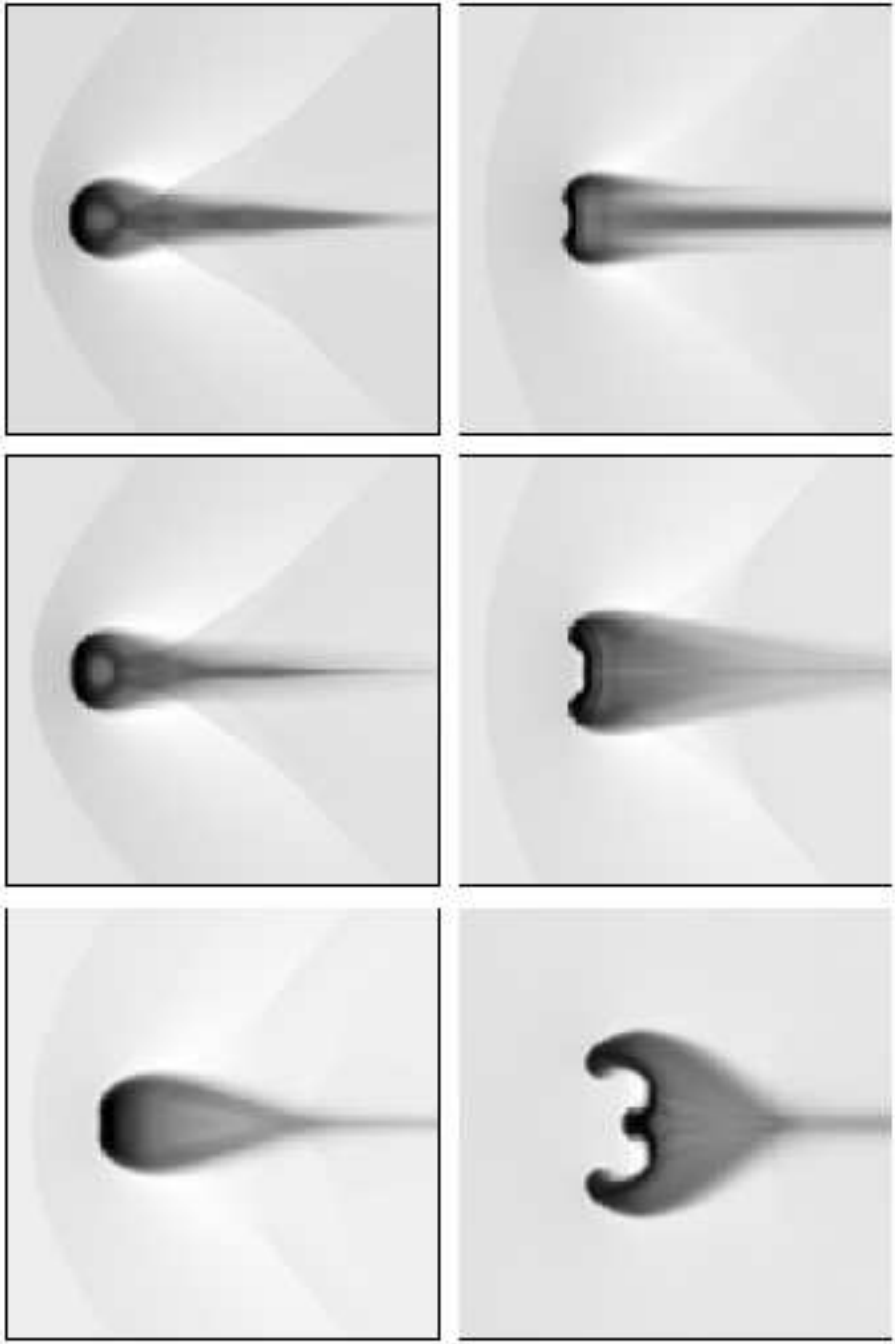
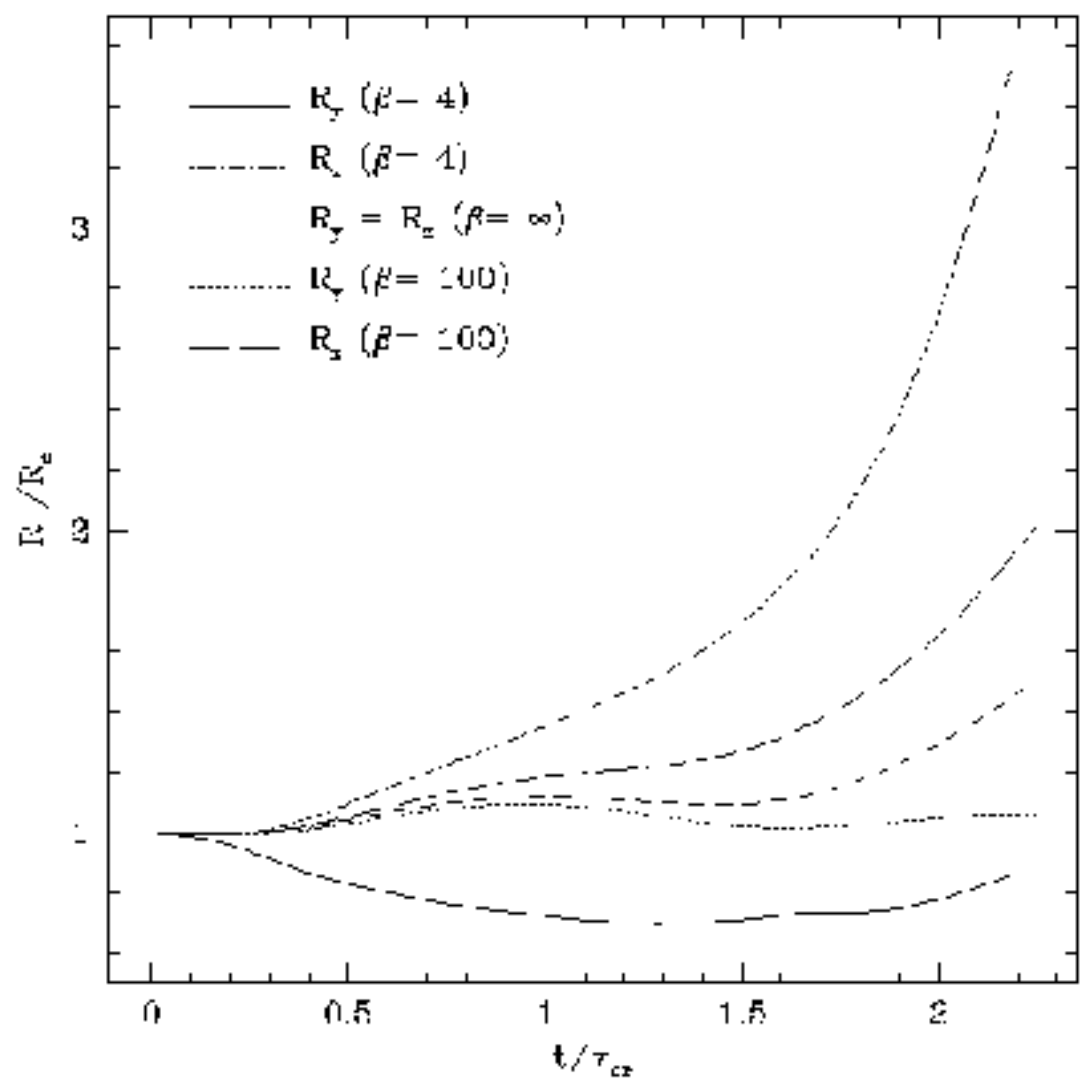
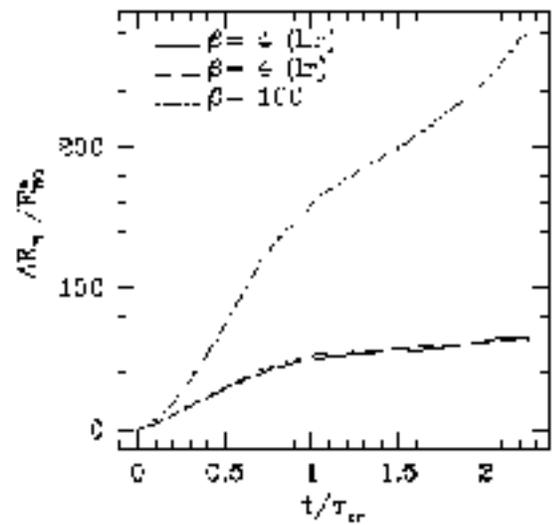
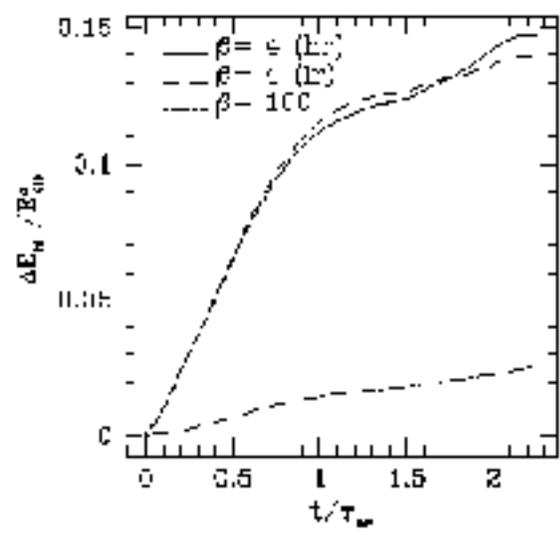
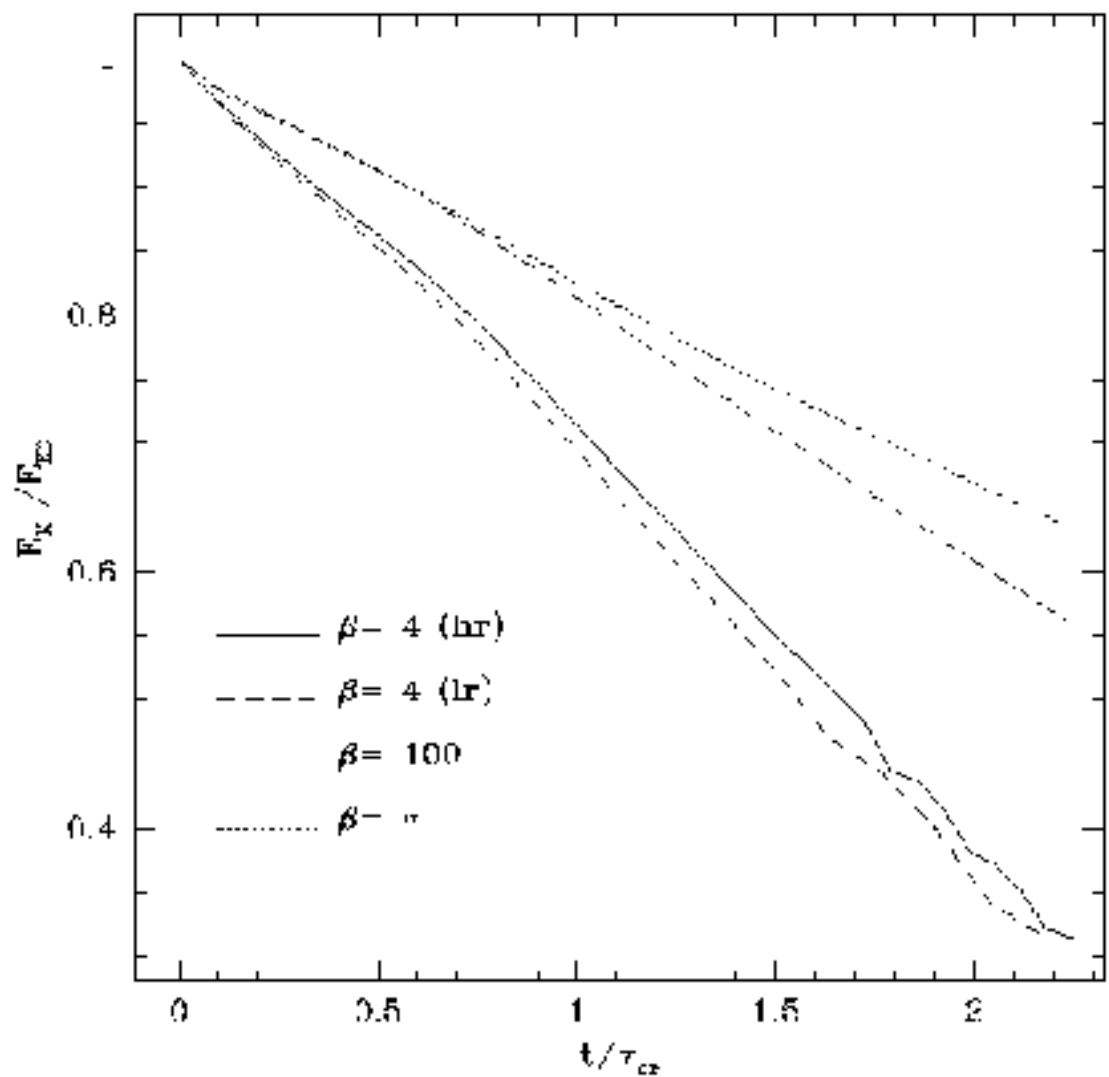
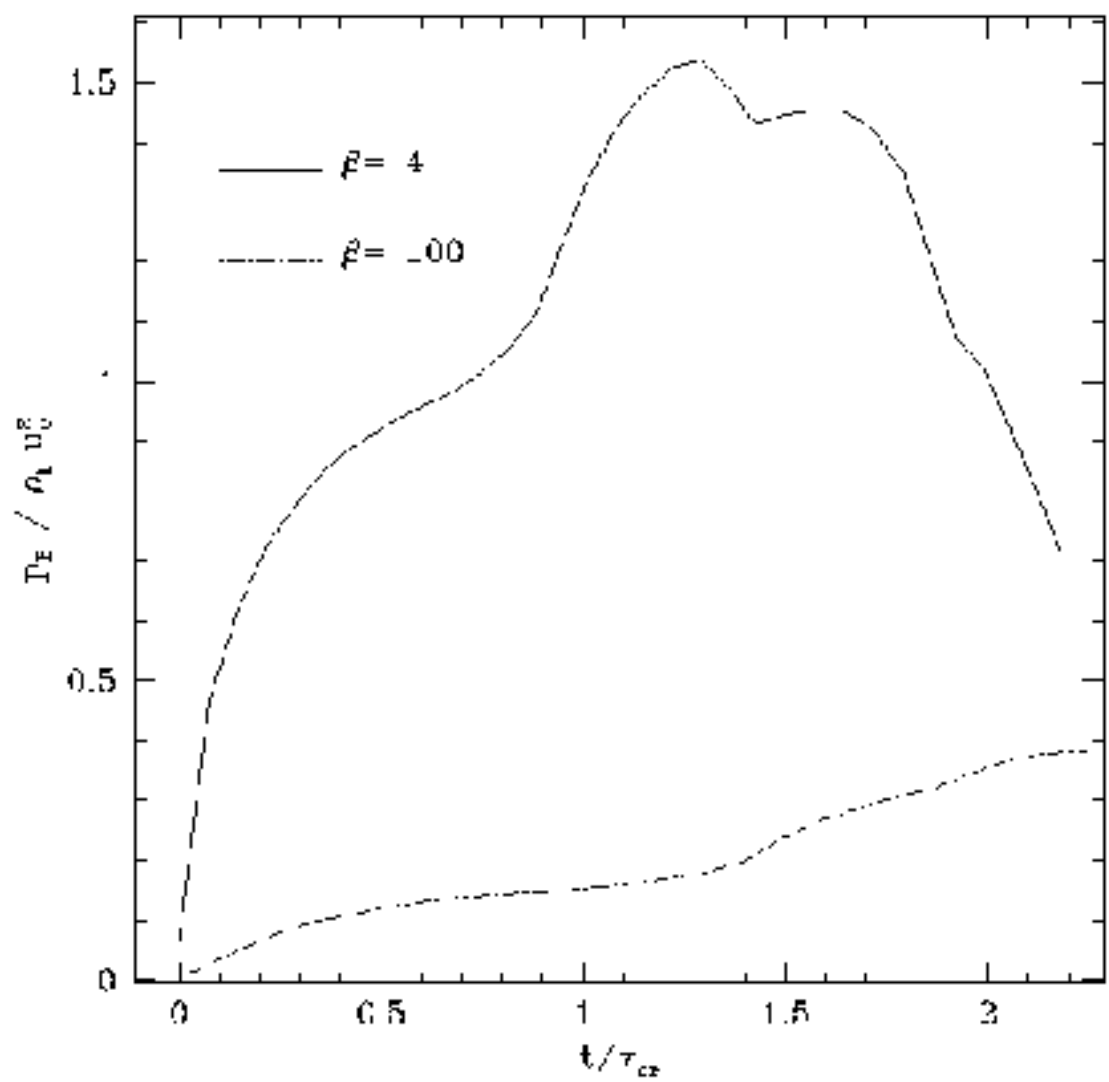
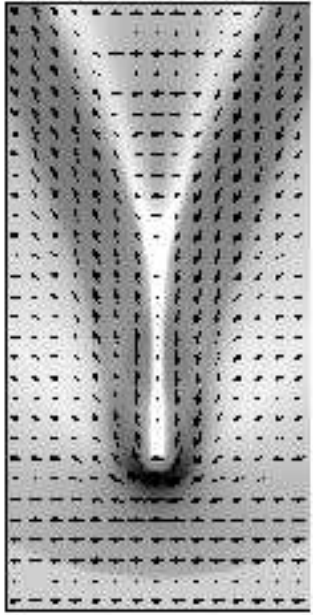
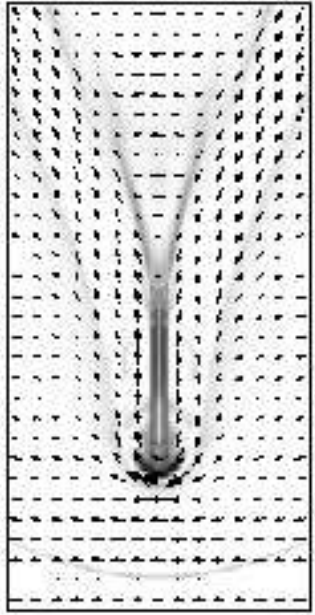
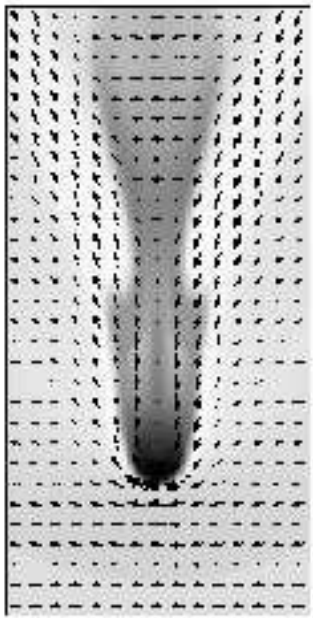
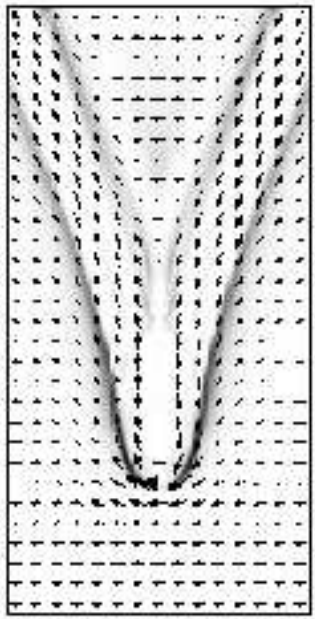
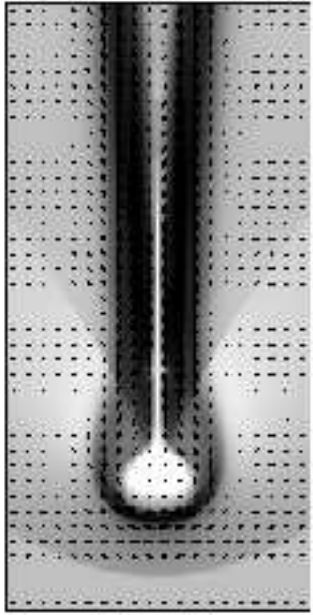
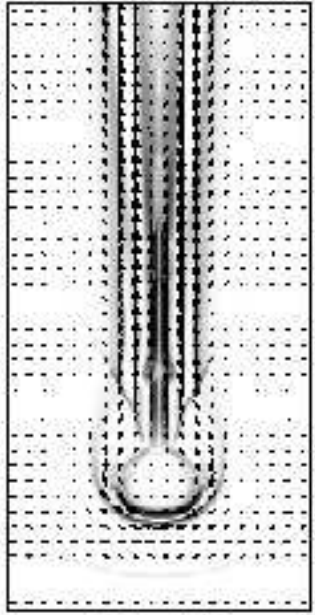
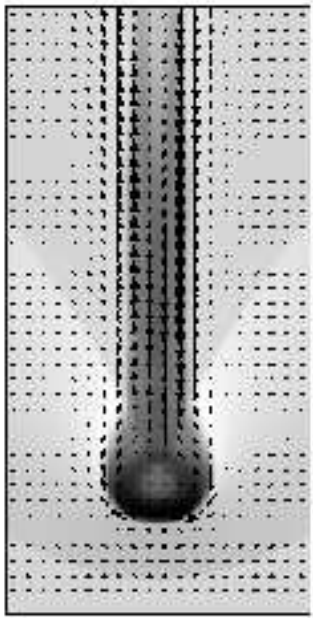
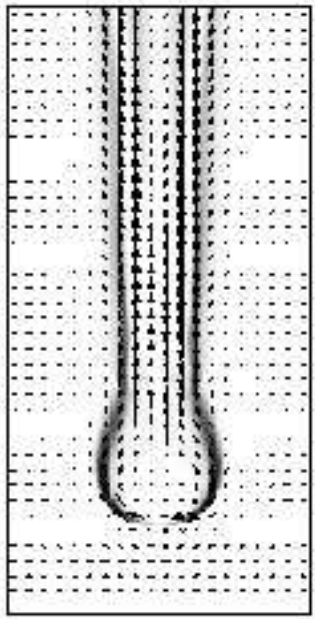
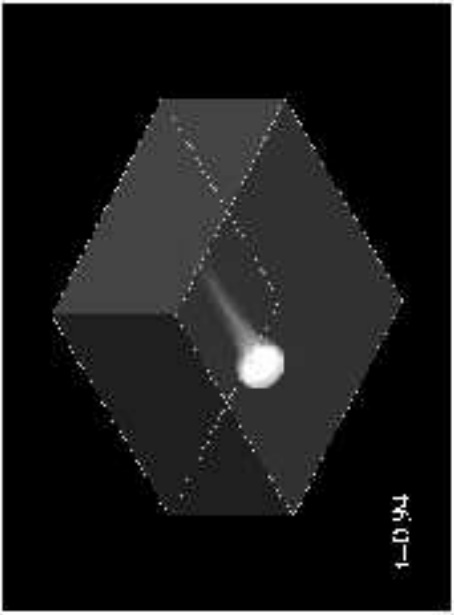
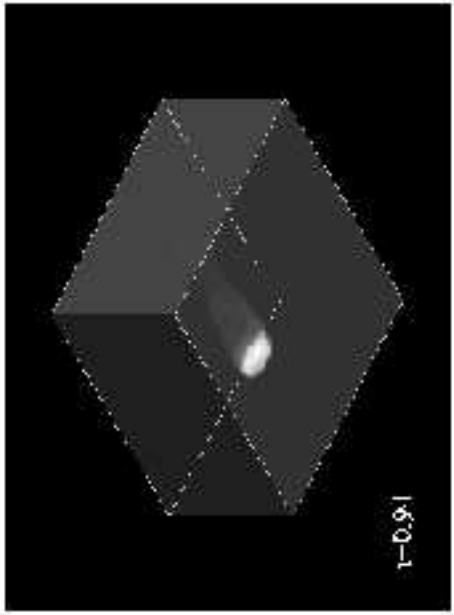
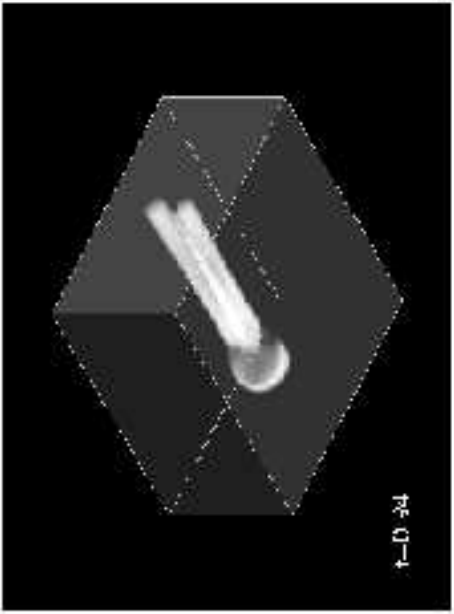
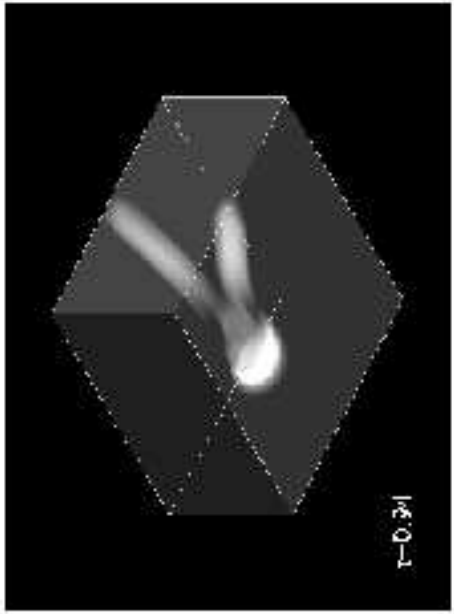
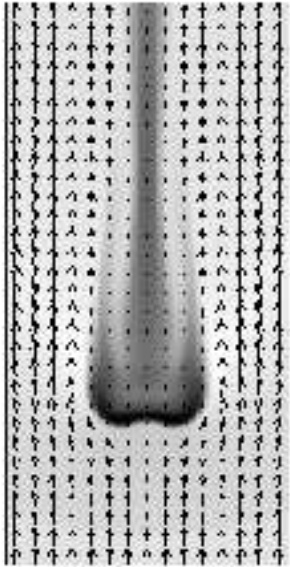
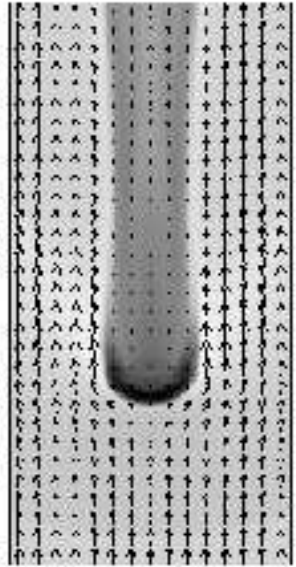
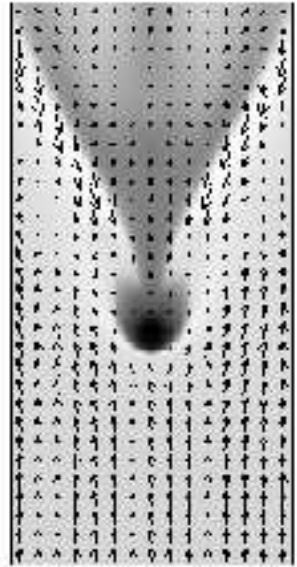
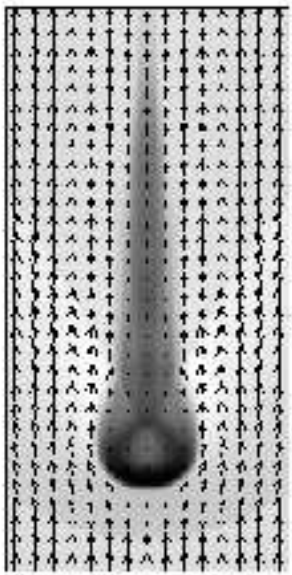
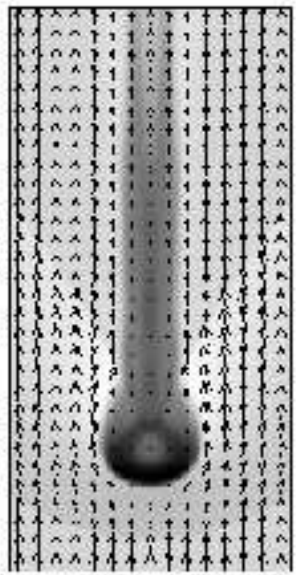
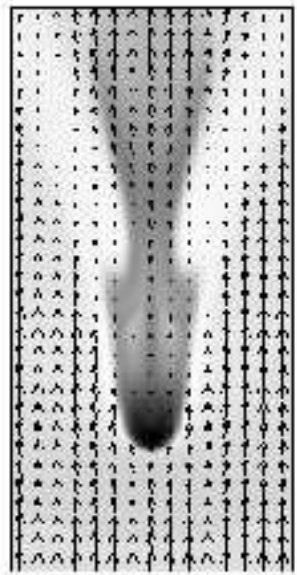
\end{document}